\documentclass[showpacs,showkeys,amsmath,amssymb,aps,twocolumn,pre]{revtex4-1}
\usepackage{graphicx}
\usepackage{dcolumn}
\usepackage{bm}
\begin{document}

\preprint{}

\title{Subdiffusive rocking ratchets in viscoelastic media: 
transport optimization and thermodynamic 
efficiency in overdamped regime}

\author{Vasyl O. Kharchenko}
 \email{vasiliy@ipfcentr.sumy.ua}

\affiliation{Institute of Applied Physics, 58 Petropavlovskaya str., 40030 Sumy, Ukraine and 
Institute for Physics and Astronomy, University of Potsdam, 
Karl-Liebknecht-Str. 24/25,
14476 Potsdam-Golm, Germany }

\author{Igor Goychuk}
 \email{igoychuk@uni-potsdam.de, corresponding author}
 
\affiliation{Institute for Physics and Astronomy, University of Potsdam, 
Karl-Liebknecht-Str. 24/25,
14476 Potsdam-Golm, Germany}

\date{\today}

\begin{abstract}

We study subdiffusive overdamped Brownian ratchets periodically rocked by an external 
zero-mean force in viscoelastic media 
within the framework of non-Markovian
Generalized Langevin equation (GLE) approach and associated multi-dimensional Markovian 
embedding dynamics. 
Viscoelastic deformations of the medium caused by the transport particle are modeled 
by a set of auxiliary Brownian quasi-particles elastically coupled 
to the transport particle and characterized by a hierarchy of relaxation times which obey
a fractal scaling. 
The most slowly relaxing deformations which cannot immediately follow
to the moving particle imprint long-range memory about its previous positions 
and cause subdiffusion and anomalous transport on a 
sufficiently long time scale. This anomalous behavior 
is combined with normal diffusion and transport on an initial time scale of overdamped motion. 
Anomalously slow directed transport in a periodic ratchet potential with broken space inversion symmetry
emerges due to a violation of the thermal detailed balance 
by a zero-mean periodic driving and is optimized with frequency of driving, its amplitude,
and temperature. Such optimized anomalous transport can be  low dispersive and 
characterized by a large generalized
Peclet number. Moreover, we show 
that overdamped subdiffusive ratchets can sustain a substantial load and do a useful work.
The corresponding thermodynamic efficiency decays
algebraically in time since the useful work done against a load scales sublinearly with time 
following to the transport particle position, but the energy pumped by an external force scales 
with time linearly.
Nevertheless, it  
can be transiently appreciably high and 
compare well with the thermodynamical
efficiency of the normal diffusion 
overdamped ratchets on sufficiently long temporal and spatial scales.

\end{abstract}
\maketitle

\section{Introduction}

Brownian motors can be described as prototype models for the bio-molecular motors in biological cells 
that work out of
thermal equilibrium \cite{JulicherRev,ReimannRev,MarchesoniHanggi}. 
Most research  in this area 
is devoted to 
normal classical ratchet transport, where both the mean position and the position variance 
grow linearly in time, i.e. $\langle\delta x(t) \rangle\propto t $ and 
$\langle\delta x^2(t) \rangle\propto t$,
respectively. 
At the same time, anomalous diffusion and transport \cite{Bouchaud,Hughes,Metzler} 
become increasingly popular 
as indicated by 
huge literature produced during the last 
20 years (see e.g. recent overview papers in \cite{WSBook} and Fig. 1 in 
Ref. \cite{WSBookG} therein). 
 Experimental works discover that both viscoelasticity  of dense polymer solutions 
 \cite{Doi,Amblard,Waigh} and colloids \cite{Russel,Mason} such 
 as e.g. cytosol of biological cells \cite{Caspi,Guigas,Mizuno,Holek,Wilhelm,Weber}, 
 and spatial 
 inhomogeneity and disorder 
 \cite{BarkaiPT,Weigel} in such media 
 can entail subdiffusion, $\langle\delta x^2(t) \rangle\propto t^\alpha$, with
a power law exponent $0<\alpha<1$. This naturally inspires the question on  how can natural 
molecular motors  operate
in such crowded environments featured by a large macroscopic (zero-frequency) viscosity which causes
subdiffusion of macromolecules on a transient mesoscopic spatial (several micrometers) and time (up to
several minutes) scales \cite{Guigas,GoychukFNL,GoychukPRE12}. In fact, such transient 
transport phenomena are faster on mesoscopic scales than one expects \cite{GoychukFNL,GoychukPRE12} for
such highly viscous media from their effective macroscopic viscosity coefficient which depends
on the Brownian particle size \cite{NelsonBook,Luby}.
To clarify the problem requires first to generalize  well known toy Brownian ratchets models 
such as rocking ratchet
or flashing ratchet towards anomalous viscoelastic dynamics with memory. The first related steps were done
recently in Refs. \cite{G10,GKh12} for rocking ratchets, and in Ref. \cite{NJP12} for flashing ratchets
within the framework of nonlinear Generalized Langevin Equation (GLE) 
approach \cite{Kubo,Zwanzig} applied to 
viscoelastic stochastic dynamics 
with memory within a generalized Maxwell-Langevin Markovian multi-dimensional embedding dynamics \cite{G09,G12}.  
In particular, in Ref. \cite{GKh12} it has been shown that such viscoelastic rocking ratchets are genuine 
ratchets capable
to sustain a sufficient load in the direction opposite to rectified motion and perform thus a useful work. 
Moreover, an optimal subdiffusive ratchet transport
reflects synchronization between the potential periodic tilts and advancing the Brownian particle over one or two 
spatial substrate periods
in the transport direction. Such a synchronization can be interpreted as 
stochastic resonance occurring in a highly non-Markovian dynamics 
upon a thermal noise intensity variation \cite{GKh12}. Furthermore, 
a synchronization between the potential periodic flashes and nonlinear oscillations 
within the potential wells is responsible
for an optimization of the ratchet transport in the case of flashing ratchets \cite{NJP12}. 
Clearly,  similar features are simply impossible within
an alternative subdiffusive transport mechanism based on continuous time random walks (CTRWs) with divergent 
mean residence times in 
traps or associated fractional Fokker-Planck dynamics (FFPD) \cite{Metzler}. Moreover, such anomalous transport 
can be low dispersive, with 
large or even diverging (for vanishingly small temperature, cf. in \cite{NJP12}) generalized Peclet 
number, which presents the ratio of scaled sub-velocity and sub-diffusion coefficient. 
This is in a sharp
contrast with highly dispersive CTRW and FFPD transport featured by a vanishing generalized 
Peclet number \cite{G12}.
In spite of these recent advances, many important fundamental questions remain but open.
In particular, in our previous works \cite{G10,GKh12,NJP12}
the inertial
effects in the anomalous ratchet transport were very essential. Will the anomalous ratchet transport persist 
also in the
overdamped limit, where the inertial effects are entirely neglected? This is the first important question
which is answered in affirmative with this work.  The second important question concerns 
thermodynamic efficiency of such anomalous isothermal anomalous Brownian motors. Namely, which portion 
of the energy provided by an external field for transport can be transformed into a useful work against a load? 
Do the conventional notion of power, i.e. the work done per unit of time, remains meaningful for anomalous
transport, or it requires a generalization? 
This is one of fundamental questions which we address in this work for anomalous ratchet transport.

In this article, we study subdiffusive \textit{overdamped}  Brownian motors operating in viscoelastic media
and periodically rocked by an external 
force. It will be shown that 
subdiffusive current can be optimized with the driving frequency and temperature also in the
overdamped limit. Such anomalous ratchets can be characterized by 
a very good transport quality (coherence) at sufficiently low temperatures in spite of the anomalous
character of  transport. 
 Studying dependence of subvelocity 
on load we will show that the considered Brownian ratchet is a genuine one. It does a useful work
against a load and
we find the corresponding thermodynamic
efficiency which turns out to be a slowly decaying function of time. This is because the energy pumped
into rectified motion scales linearly with time, but the subdiffusive transport is sublinear.
Nevertheless, this efficiency is not vanishingly small and it compares well with the efficiency of
the corresponding normal diffusion ratchets on the time scale of simulations, which is a surprise.

\section{Model}

Let us consider the following model of anomalous Brownian motion. A Brownian particle moving with velocity 
$dx(t)/dt\equiv \dot x(t)$ in a dense water solution of polymers (e.g. cytosol of biological
cells) experiences
Stokes memoryless friction with friction coefficient $\eta_0$ (corresponding to water) and, in addition, 
a frequency-dependent friction with memory
which is characterized  by a fractional friction coefficient $\eta_{\alpha}$. It corresponds
to a polymeric fluid. Considering a general case of linear friction
with memory, $f_{\rm mem}(t)=-\int_0^{t}\eta(t-t')\dot x(t')dt'$, for a particle starting its motion at $t_0=0$, 
the normal contribution
corresponds to the memory kernel $\eta(t)=2\eta_0\delta(t)$ and frictional force, $f_{\rm Stokes}(t)=-\eta_0 dx(t)/dt$, 
whilst an anomalous one emerges for 
$\eta(t)=\eta_{\alpha}t^{-\alpha}/\Gamma(1-\alpha)$, where $0<\alpha<1$ and $\Gamma(\cdot)$ is gamma-function. 
The corresponding term with memory, which captures e.g. viscoelastic effects, can be abbreviated as  
$f_{\alpha}(t)=-\eta_\alpha 
d^{\alpha}x(t)/dt^{\alpha}$ using the notion of fractional Caputo derivative, just per its definition
\cite{Gorenflo}.
In accordance with the second 
fluctuation-dissipation theorem (FDT) \cite{Kubo}, these dissipative forces are complemented by the
corresponding mutually 
independent thermal fluctuation forces $\xi_0(t)$ and $\xi_{\alpha}(t)$, which are Gaussian, zero-mean, and completely
characterized by the autocorrelation functions $\langle\xi_0(t)\xi_0(t')\rangle=2k_B T\eta_0\delta(t-t')$ and
\begin{eqnarray}\label{FDT}
\langle\xi_\alpha(t)\xi_\alpha(t')\rangle=k_B T\eta_\alpha | t-t'|^{-\alpha}/\Gamma(1-\alpha),
\end{eqnarray}
at the environmental temperature $T$. In the presence of external force field $f(x,t)=-\partial V(x,t)/\partial x$ 
the motion of an
overdamped Brownian particle is described by an overdamped Generalized Langevin Equation (GLE) 
\cite{Kubo,Zwanzig}:  
\begin{equation}
\eta_0\frac{{\rm d} x}{{\rm d}t}+\eta_\alpha\frac{{\rm d}^\alpha x}{{\rm d}t^\alpha}=f(x,t)+
 \xi_0(t)+\xi_{\alpha}(t).
\label{GLE1}
\end{equation}
Furthermore, we shall consider a 
spatially asymmetric periodic ratchet potential \cite{Bartussek} 
\begin{equation}
 U(x)=-U_0\left[\sin\left(\frac{2\pi x}{L}\right)+\frac{1}{4} \sin\left(\frac{4\pi x}{L}\right) \right] 
 \label{potential}
\end{equation}
with amplitude $U_0$ and spatial 
period $L$. The motion is  driven also 
by an external periodic force $f_{\rm ext}(t)=A\cos(\Omega t)$ with amplitude 
$A$ and driving frequency $\Omega$. The FDT (\ref{FDT}) ensures so that the energy dissipated is always 
balanced by the energy gained from the
environment at the thermal equilibrium ($f_{\rm ext}\to 0$) so that there is no net heat exchange and 
the kinetic degree of freedom has energy
$k_BT/2$ on average. External force $f_{\rm ext}$ is expected to violate  thermal detailed balance and cause 
a net directed motion of the Brownian
particles beyond thermal equilibrium.
This net motion can be directed against a loading in the opposite direction constant 
force $f_0$ and do a useful
work against such a load. Altogether,  $V(x,t)=U(x)-f_{\rm ext}(t)x+f_0x$.

Let us scale further
the coordinate $x$ in the units of $L$ and time $t$ in the units of 
$\tau_r=\left(4\pi^2U_0/L^2\eta_\alpha\right)^{-1/\alpha}$.
In these units, the GLE (\ref{GLE1}) reads
\begin{eqnarray}
 \eta_0\frac{{\rm d}x}{{\rm d}t}+\frac{{\rm d}^\alpha x}{{\rm d}t^\alpha}=
 \frac{1}{2\pi}\Big[f(x,t)+\sqrt{2T\eta_0}\zeta_0(t)\nonumber \\
 +\sqrt{T/\Gamma(1-\alpha)}\zeta_{\alpha}(t)\Big],
\label{GLE2}
\end{eqnarray}
where the friction coefficient $\eta_0$ is scaled in the units of $\eta_\alpha \tau_r^{1-\alpha}$, 
temperature $T$ in the units of $U_0/k_B$ and 
\begin{eqnarray}
f(x)=\cos(2\pi x)+(1/2)\cos(4\pi x)\nonumber \\ +A\cos(\Omega t)-f_0,
\end{eqnarray}
where $A$ and $f_0$ are scaled in the units of $2\pi U_0/L$.
Moreover,  $\langle\zeta_0(t)\zeta_0(t')\rangle=\delta(t-t')$,
and  $\langle\zeta_\alpha(t)\zeta_\alpha(t')\rangle=1/|t-t'|^{\alpha}$ in these units.
 Next, we follow to the road of Markovian embedding in Refs.  \cite{G09,G12} and approximate
the memory kernel by a sum of exponentials,  
$\eta(t)=\sum_{i=1}^{N}k_i\exp(-\nu_it)$ and the corresponding noise $\xi_{\alpha}(t)$ by a sum of Ornstein-Uhlenbeck
processes. By choosing the spectrum of relaxation rates scaled as  $\nu_i=\nu_0/b^{i-1}$ via a maximal 
relaxation rate $\nu_0$ and a scaling parameter $b>1$ the corresponding power dependence $t^{-\alpha}$ can be 
nicely approximated over 
about $r=N\log_{10}b-2$ time decades between two time 
cutoffs $\tau_{l}=\nu_0^{-1}$ and $\tau_{h}=b^{N-1} \tau_{l}$. To ensure a power law scaling
one chooses $k_i\propto \nu_i^{\alpha}$, or 
 $k_i=C_{\alpha}(b)\nu_i^{\alpha}/\Gamma(1-\alpha)$, where $C_{\alpha}(b)$ is a numerical fitting constant
which mostly depends on $\alpha$ and $b$ for a sufficiently large $r$ and $N$. 
Weak dependences of $r$ on $N$ and $b$
 ensure a very powerful numerical
approach to integrate fractional stochastic 
non-Markovian dynamics with a well-controlled numerical accuracy (of several percents
in this work). Alternatively, it can be considered as an independent approach to anomalous
transport which is even not bounded by a strict requirement on the power scaling.
Approximation of the memory kernel by a sum of exponentials can be derived from an
experiment, see a practical example in Fig. 3 in Ref. \cite{G12}, where the sum of just four
exponentials suffices to fit a power law memory kernel extending over four time decades.
This approach allows also for a vivid physical interpretation
in terms of  viscoelastic forces  $u_i=-k_i(x-x_i)$ caused by overdamped Brownian particles modeling viscoelastic
degrees of freedom of the environment and corresponding to
viscoelastic deformations of the medium (principal modes). 
These auxiliary quasi-particles are elastically attached to the transport particle
with spring constants $k_i$ and subjected to Stokes frictional forces with frictional constants 
$\eta_i=k_i/\nu_i$.  This leads to the following Markovian embedding dynamics with uncorrelated white
noise sources,
$\langle\zeta_i(t)\zeta_j(t')\rangle=\delta_{ij}\delta(t-t')$:
\begin{equation}
\begin{split}
  &\eta_0\dot{x}=\frac{1}{2\pi}f(x,t)-\sum\limits_{i=1}^Nk_i(x-x_i)+\frac{\sqrt{2T\eta_0}}{2\pi}\zeta_0(t),\\
  &\eta_i\dot{x}_i=k_i(x-x_i)+\frac{1}{2\pi}\sqrt{2T\eta_i}\zeta_i(t).
\end{split}
\label{GLEf}
\end{equation}
Initial $x_i(0)$ must be thermally (Gaussian) distributed around $x(0)$, with $\langle [x_i(0)-x(0)]^2\rangle=T/k_i$,
in order to have complete equivalence with above GLE description for a memory kernel being a sum
of exponentials, which is, of course, an approximation to the considered power-law memory kernel \cite{G12}. 
The accuracy of this approximation is, however, well controlled.
We do the corresponding sampling of $x_i(0)$ below. Otherwise, there
would be a transient force present in GLE reflecting thermal equilibration of the medium disturbed initially
by the Brownian particle (e.g. upon its insertion), 
or corresponding aging
effects \cite{G12}. The general presence of such transients adds a flexibility to this modeling approach.
However, we shall not consider such transient aging effects because we are interested in an asymptotic transport
regime, which is not influenced by initial transients.
Thereafter we fix $\alpha=1/2$ and choose $b=10$, $C_{1/2}(10)=1.3$ and $\nu_0=100$. 
Most presented below results were obtained using an ensemble averaging over $10^4$ trajectories.
Simulations are done with the help of the stochastic Heun method \cite{Heun} on the graphical processor units (GPUs) 
with double precision 
\cite{Footnote}.
This technique provided an effective acceleration of numerics by a factor of about 100 for the studied system 
over the standard CPUs computing on modern commodity processors. 
The total integration time of GLE (\ref{GLEf}) was varied in the interval $t_{total}\in[3\times10^5\ldots 10^6]$ 
and, the time step was $\Delta t=2\times 10^{-3}$. The number of auxiliary particles was fixed to 
$N=12$.

As discussed previously  \cite{G09,G12,NJP12}, $N$ auxiliary particles can be  
roughly divided into the groups of fast, $N_f$, and slow $N_s=N-N_f$ particles. 
This division can be made upon comparison of the mean time of transitions made by central Brownian 
particle to the neighboring potential well with the relaxation times $\nu_i^{-1}$ of the corresponding 
viscoelastic force components $u_i$. 
Fast medium's deformations move together with the Brownian particle, forming together 
a quasi-particle. It reminds polaron in condensed matter physics, i. e. a naked particle 
plus deformation of its
nearest environment which are considered together as a compound particle. 
Mean viscoelastic force created by such a nearest environment equals to zero on average on the time
scale of slow motion. 
However, the most sluggish deformations temporally 
imprint the medium's memory
about the previous particle's positions and create a quasi-elastic slow varying retarding force 
acting in the direction opposite to the transport direction. These viscoelastic deformations
 introduce long-lived negative correlations
in the particle displacements. Such a mechanism leads to anomalously slow diffusion and  transport 
with $\langle\delta x^2(t)\rangle \propto t^\alpha$ and $\langle x(t)\rangle \propto t^\alpha$, 
respectively, where $\langle\delta x^2(t)\rangle=\langle x^2(t)\rangle-\langle x(t)\rangle^2$. 
The corresponding subdiffusion coefficient $D_\alpha$ and the subvelocity $v_\alpha$ are defined as follows:
\begin{equation}
 \begin{split}
  &D_\alpha=\frac{1}{2}\Gamma(1+\alpha)\lim\limits_{t\to\infty}\frac{\langle\delta x^2(t)\rangle}{t^\alpha},\\
  &v_\alpha=\Gamma(1+\alpha)\lim\limits_{t\to\infty}\frac{\langle x(t)\rangle}{t^\alpha}.
 \end{split}
\label{vada}
\end{equation}
The limit should be understood as a physical limit in the following 
sense: $t$ is large but yet much smaller than the time cutoff
$\tau_0 b^{N-1}$. The latter time scale is made not attainable (and thus irrelevant) in our simulations. 
To characterize the coherence and the quality of the transport we shall use a Generalized Peclet 
number $Pe_\alpha=v_\alpha L/D_\alpha$ \cite{G10}. Such a Peclet number is a natural quantifier for
the coherence quality of stochastic transport in periodic potentials.  It measures the ratio of the 
mean traveling distance $\langle x(t)\rangle $ (in units of $L$) to the diffusional 
spread $\langle \delta x^2(t)\rangle $ (in units of $L^2$)
\cite{Lindner}. 

\section{Results and discussion}

\begin{figure}
\resizebox{0.75\columnwidth}{!}{\includegraphics{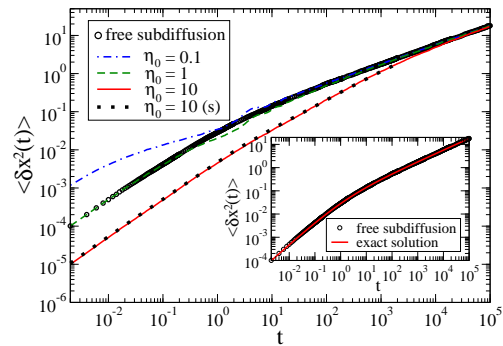}}
\caption{(Color online) Subdiffusion in driven ratchet potential for various $\eta_0$,
 $\Omega=1.0$, $A=1.0$, and $T=1.0$: 
solid, dash and dash-dot curves correspond to ensemble averaging; dot curve (s) relates to a 
single trajectory averaging in Eq. (\ref{time-av}) and the agreement indicates ergodicity. The insert demonstrates 
an excellent agreement between the analytical result in Eqs. (\ref{exact}), (\ref{exact2}) and numerics
for free diffusion case
over eight time decades for $\eta_0=1$.}
\label{Fig1}       
\end{figure}

First, we tested numerics and compared the results for the ensemble-averaged position variance $\langle\delta x^2(t)\rangle$ 
with the exact result
 in the absence of potential which can be readily obtained from a general expression for the
considered particular case of GLE, see in \cite{G12}, using the Laplace transform of generalized
Mittag-Leffler function from Ref. \cite{Haubold}. The result reads 
(in the original nonscaled units),
\begin{eqnarray}\label{exact}
\langle\delta x^2(t)\rangle =2D_{0} t E_{1-\alpha,2}[-(t/\tau_0)^{1-\alpha}],
\end{eqnarray}
where $E_{a,b}(z):=\sum_0^{\infty}z^n/\Gamma(an+b)$ is generalized Mittag-Leffler function,
$D_0=k_BT/\eta_{0}$ is a normal diffusion coefficient, and $\tau_0=(\eta_0/\eta_{\alpha})^{1/(1-\alpha)}$
is a transient time constant. 
For small argument, $E_{a,b}(z\ll 1)\approx 1$, and for large argument,
$E_{a,b}(z\gg 1)\sim -1/[\Gamma(b-a)z]$, in the leading order of $z^{-1}$. 
Hence, $E_{1-\alpha,2}(-z^{1-\alpha})\sim z^{\alpha-1}/\Gamma(1+\alpha)$, for $z\gg 1$.
This result shows that at small times, $t\ll \tau_0$,
the diffusion is normal, $\langle\delta x^2(t)\rangle\approx 2D_0 t$, whereas at large times, 
$t\gg \tau_0$, it becomes anomalously slow, 
$\langle\delta x^2(t)\rangle\approx 2D_{\alpha} t^{\alpha}/\Gamma(1+\alpha)$, with $D_{\alpha}=k_BT/\eta_{\alpha}$. 
$\tau_0$ defines a characteristic time separating these two different regimes. Notice that it scales as 
$\eta_0^{1/(1-\alpha)}$ with $\eta_0$. 
For $\alpha=1/2$, this general result can be expressed in terms of complementary error function and 
a power law dependence,
\begin{eqnarray}\label{exact2}
\langle\delta x^2(t)\rangle && = 2D_{\alpha}\left \{2\sqrt{\frac{t}{\pi}} \right. \\&&+ \left. \sqrt{\tau_0}
\left [ e^{t/ \tau_0}{\rm erfc}\left (\sqrt{\frac{t}{\tau_0}}\right ) -1\right ]
 \right\} \nonumber\;.
\end{eqnarray}
In the used scaling, it is a solution of Eq. (\ref{GLE2}), with $f\to 0$, 
$\eta_{\alpha}\to 1$, and $D_{\alpha}\to 1/(2\pi)^2$, at $T=1$. 
The agreement with numerics is excellent, cf. insert
in Fig. \ref{Fig1}.

\begin{figure}
\resizebox{0.75\columnwidth}{!}{\includegraphics{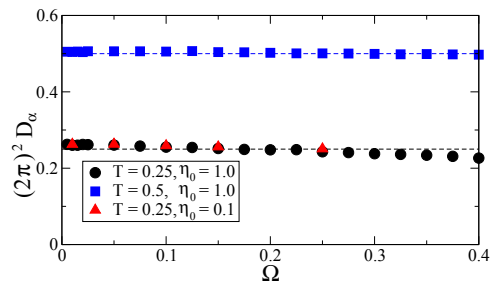}}
\caption{(Color online) Scaled subdiffusion coefficient $D_\alpha$ 
as function of driving frequency $\Omega$ for several different values of 
temperature $T$ and friction coefficient $\eta_0$.}
\label{Fig2}       
\end{figure}

It has been shown previously for subdiffusive dynamics with inertial effects \cite{G09,G10,G12,WSBookG} that 
a periodic potential does not influence the variance $\langle\delta x^2(t)\rangle$ and
the corresponding subdiffusion coefficient $D_\alpha$ in the asymptotical limit. This was named 
a universality class of viscoelastic subdiffusion in tilted periodic potentials in Refs. \cite{G12,WSBookG}.
This remarkable property is also valid for the overdamped ratchets considered in this work.
The explanation is also similar: most sluggish viscoelastic modes of the environment determine
the asymptotic character of subdiffusion not being affected by external static fields. 
Strong time-periodic fields can make some influence on viscoelastic subdiffusion pumping
energy into the system at some rate. However,
this influence was not strong  even in the presence of inertial effects \cite{G09}, as Figs. \ref{Fig1}
and \ref{Fig2}
also illustrate for the inertia-free dynamics. For a sufficiently large driving frequency $\Omega$
some deviation from the force-free subdiffusion coefficient, $D_\alpha^{(0)}=T/(2\pi)^2$, occurs
in Fig. \ref{Fig2}. However, it is not appreciably strong. 

One can see in Fig. \ref{Fig1} that an increase in viscous friction $\eta_0$ leads to extension of the initial normal
diffusion regime. 
In contrast to the case studied earlier \cite{G09,G10,G12,GKh12} there is no ballistic regime here due the absence of 
inertial effects.
Initially diffusion is normal, $\langle\delta x^2(t)\rangle\sim t$, in accordance with above analysis. 
Finally, on a large time scale 
diffusion becomes anomalous, $\langle\delta x^2(t)\rangle\sim t^{1/2}$. 
Normal friction does not affect this asymptotics, leading merely to increase of the transition time $\tau_0$ 
(compare dash-dot, solid and dash curves in Fig.\ref{Fig1}). 

To check if  the 
considered dynamics is  ergodic, in accordance
with previous studies of viscoelastic subdiffusion in periodic potentials, we computed 
a time-average $\langle\delta x^2(t)\rangle_{\cal T}$ of the squared displacement,
\begin{eqnarray}\label{time-av}
 \langle \delta x^2(t)\rangle_{\cal T}=\frac{1}{{\cal T}-t}\int_{0}^{{\cal T}-t}[x(t+t')-x(t')]^2 dt',
\end{eqnarray}
over single trajectories \cite{He,Lubelski,Deng,G09}. 
Here, the total integration time $\cal T$ is chosen  much larger than the maximal time $t_{\rm max}$ for the
ensemble-averaged trajectories
$\langle \delta x^2(t)\rangle$. We used ${\cal T}/t_{\rm max}=10^3$ in our calculations.
The underlying idea of ergodicity is that the time average of a quantity, here squared 
particle displacements within a time interval of length $t$ is equal 
to a corresponding ensemble average. 
In other words, a moving time-average of the squared displacement should coincide with the ensemble-average.
Comparing solid and dotted curves in Fig.\ref{Fig1} for $\eta_0=10$ 
one can conclude that this indeed is the case. The diffusion
is clearly ergodic on the considered time scale.

\subsection{Transport dependence on driving frequency}

\begin{figure}
\resizebox{0.75\columnwidth}{!}{\includegraphics{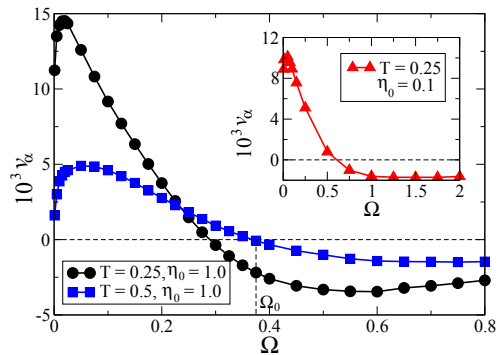}}
\caption{(Color online) Anomalous current (subvelocity $v_\alpha$)
 as function of driving frequency for different temperatures $T$ and friction coefficients $\eta_0$.}
\label{Fig3}       
\end{figure}

To compute the frequency-dependence $v_\alpha(\Omega)$ we have chosen several different values of temperature $T$
and normal friction coefficient $\eta_0$, 
and 
varied the driving frequency $\Omega$ in the window $(0\ldots2]$ at fixed amplitude $A=1$. 
The results for the anomalous current (subvelocity $v_\alpha$) as a function of driving 
frequency $\Omega$ are shown in Fig.\ref{Fig3}.

The occurrence and frequency dependence of the rectification effect is particularly interesting
in the considered overdamped dynamics. In the presence of inertial effects, rectification 
ratchet effect is suppressed
in the adiabatic frequency limit $\Omega\to 0$ \cite{G10,GKh12}. This  reflects the universality
class of anomalous GLE sub-transport in washboard potentials and is very different
from the normal diffusion case, where the rectification effect for the fluctuating tilt 
ratchets is maximal namely
in the discussed limit \cite{Bartussek,ReimannRev,MarchesoniHanggi}.
One expects that
this anomalous feature  survives also for overdamped dynamics, as Fig. \ref{Fig3} indeed confirms. 

Further, let us compare the influence of temperature and driving frequency on 
the subvelocity $v_\alpha(\Omega)$ at the fixed $\eta_0=1$ (see curves with filled circles and 
squares for small and intermediate temperatures, respectively). 
For the tested values of temperature 
the subcurrent indeed optimizes with the driving frequency. Increase in temperature leads to 
decrease in the maximal value of subvelocity and to 
shift of the optimal driving frequency towards larger values. The qualitative explanation of these effects
is similar to one for subdiffusive ratchets with inertial effects \cite{GKh12}. 
With increasing temperature the role of trapping potential
diminishes (transport is absent in the absence of ratchet potential), and the mean time to escape out 
of potential well  
decreases. We expect that the optimal frequency also corresponds to a stochastic resonance (SR) effect, 
similar to one in
Ref. \cite{GKh12}, when the mean frequency of jumps in the transport direction synchronizes 
with the corresponding potential tilts.
This question was not studied, however, in more detail for the case considered.

Furthermore, 
it has been shown for the inertial case that an increase in the driving frequency 
can lead to the subcurrent 
inversion, where the transport occurs in the counterintuitive direction. The corresponding driving frequency
is of the order of magnitude of the inverse anomalous relaxation time constant of the 
velocity autocorrelation function 
$\tau_v=(m/\eta_{\alpha})^{1/(2-\alpha)}$, in the presence of inertial effects.  
It depends on the mass of Brownian particle $m$.
We considered here, however, an overdamped limit, $m\to 0$ with $\tau_v\to 0$. For this reason, 
it is \textit{a priori}
not clear if the inversion of transport direction can occur also in the complete absence of 
inertial effects.
Numerics do reveal such an inversion for a sufficiently large frequency. 
However, the corresponding characteristic time scale is given now not by the velocity 
relaxation time constant 
$\tau_v$, but by the time scale of intrawell coordinate relaxation $\tau_r^{(eff)}$. 
Such an inversion is similar to one detected for normal diffusion
ratchets in Ref. \cite{Bartussek}.
For small $\eta_0\to 0$,
it is primarily determined by the anomalous relaxation time constant $\tau_r$, i. e.
 $\tau_r^{(eff)}\sim 
\tau_r \sim \eta_\alpha^{1/\alpha}$. 
An increase
in $\eta_0$ decelerates the intrawell relaxation process and leads to an increase in the effective relaxation time,
which becomes proportional to $\eta_0$, $\tau_r^{(eff)}\sim 
\eta_0$, in the normal diffusion limit $\eta_\alpha\to 0$. For this reason, a critical
inversion frequency, $\Omega_{\rm cr}\sim 1/\tau_r^{(eff)}$, should decrease with the increase
in $\eta_0$ at fixed $\eta_\alpha$.  Moreover, an increase in temperature should also increase the relaxation
rate leading to a larger value of $\Omega_{\rm cr}$. 
The results in Fig. \ref{Fig3} are consistent with this explanation and show the
corresponding tendencies.

\subsection{Temperature dependence of anomalous transport and its dispersion}

\begin{figure}
\resizebox{0.75\columnwidth}{!}{\includegraphics{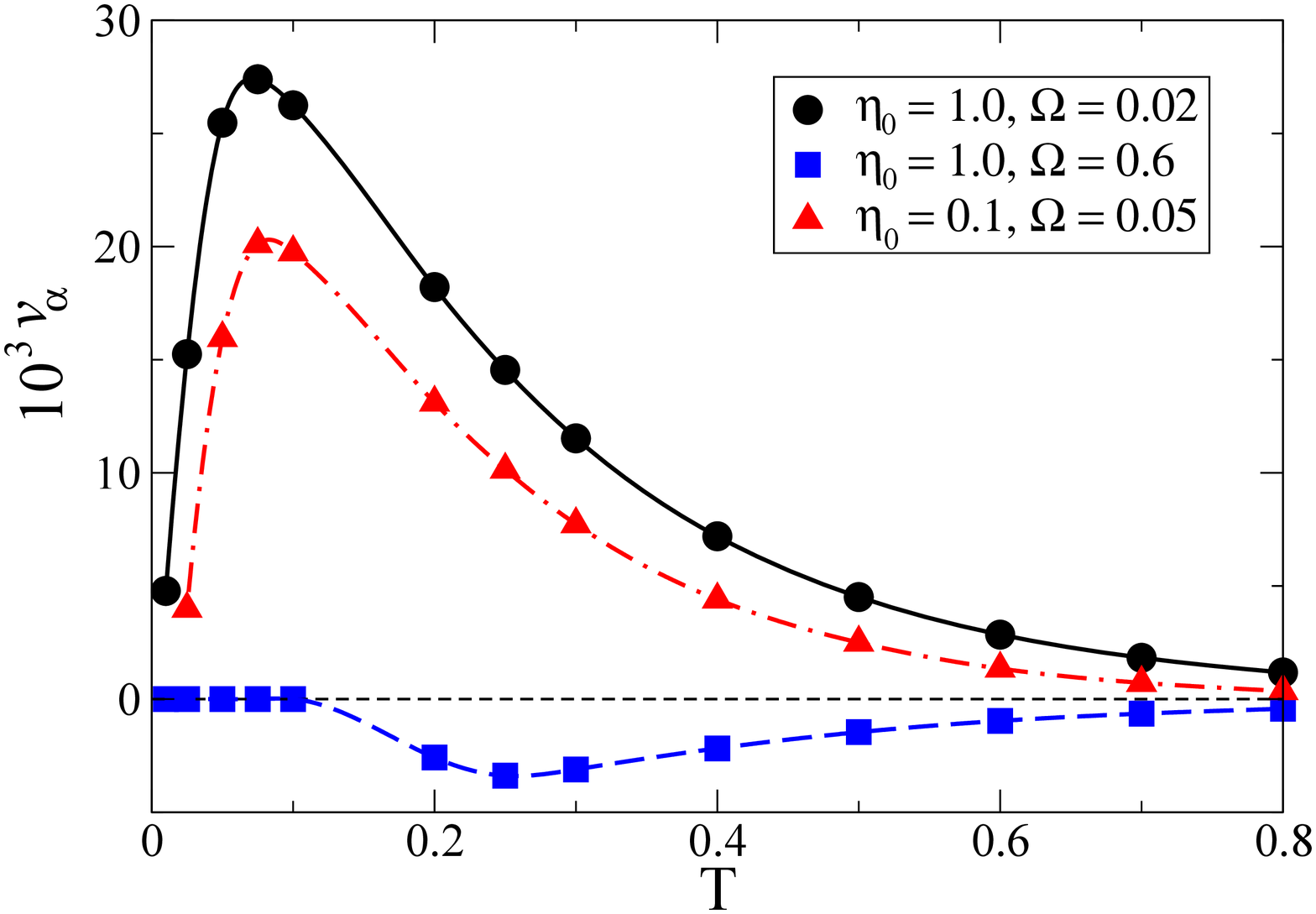}}
\caption{(Color online) Subvelocity $v_\alpha$ 
as function of temperature $T$ for several values of frequency $\Omega$ and friction coefficient $\eta_0$.}
\label{Fig4}       
\end{figure}

\begin{figure}
\resizebox{0.75\columnwidth}{!}{\includegraphics{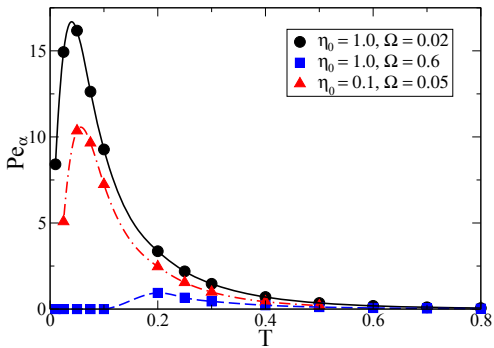}}
\caption{(Color online) Generalized Peclet number ${\rm Pe}_\alpha$
as function of temperature $T$ for several values of frequency $\Omega$ and friction coefficient $\eta_0$.}
\label{Fig5}       
\end{figure}

Given a subthreshold driving, we are dealing with a thermal noise assisted ratchet transport.
Thermal noise is necessary to overcome the potential barriers and therefore subtransport vanishes
in the limit of zero temperature, $T\to 0$. It vanishes also for a large temperature $T\gg U_0$,
when the potential ceases to matter. Therefore, an optimization with temperature is expected, also
for the inverted transport regime. Indeed numerics reveal such an optimization clearly, see in Fig. 
\ref{Fig4} for different parameters and different regimes. 
Dependency on frequency $\Omega$ for the same $\eta_0$ implies
for sufficiently small $\Omega$ (not shown) that the maximum versus temperature corresponds
to a stochastic resonance (SR), when the overbarrier jumps synchronize with the potential tilts in the
transport direction, like in the presence of 
inertial effects \cite{GKh12}. A detailed study of such a non-Markovian SR for overdamped dynamics 
was, however, not done. It is left for a separate study. The coherence quality 
of the overdamped transport, as measured by the generalized Peclet number 
${\rm Pe}_\alpha:=v_\alpha L/D_\alpha$,  can also be rather high, like in the presence of inertial
effects. Fig. \ref{Fig5} shows this clearly for $v_\alpha$ in Fig. \ref{Fig4}. This is an expected
result since $D_\alpha\propto T$. Because of this the considered non-Markovian stochastic 
coherence resonance is shifted to smaller values of optimal $T$ as compare with SR.

\begin{figure}
\resizebox{0.75\columnwidth}{!}{\includegraphics{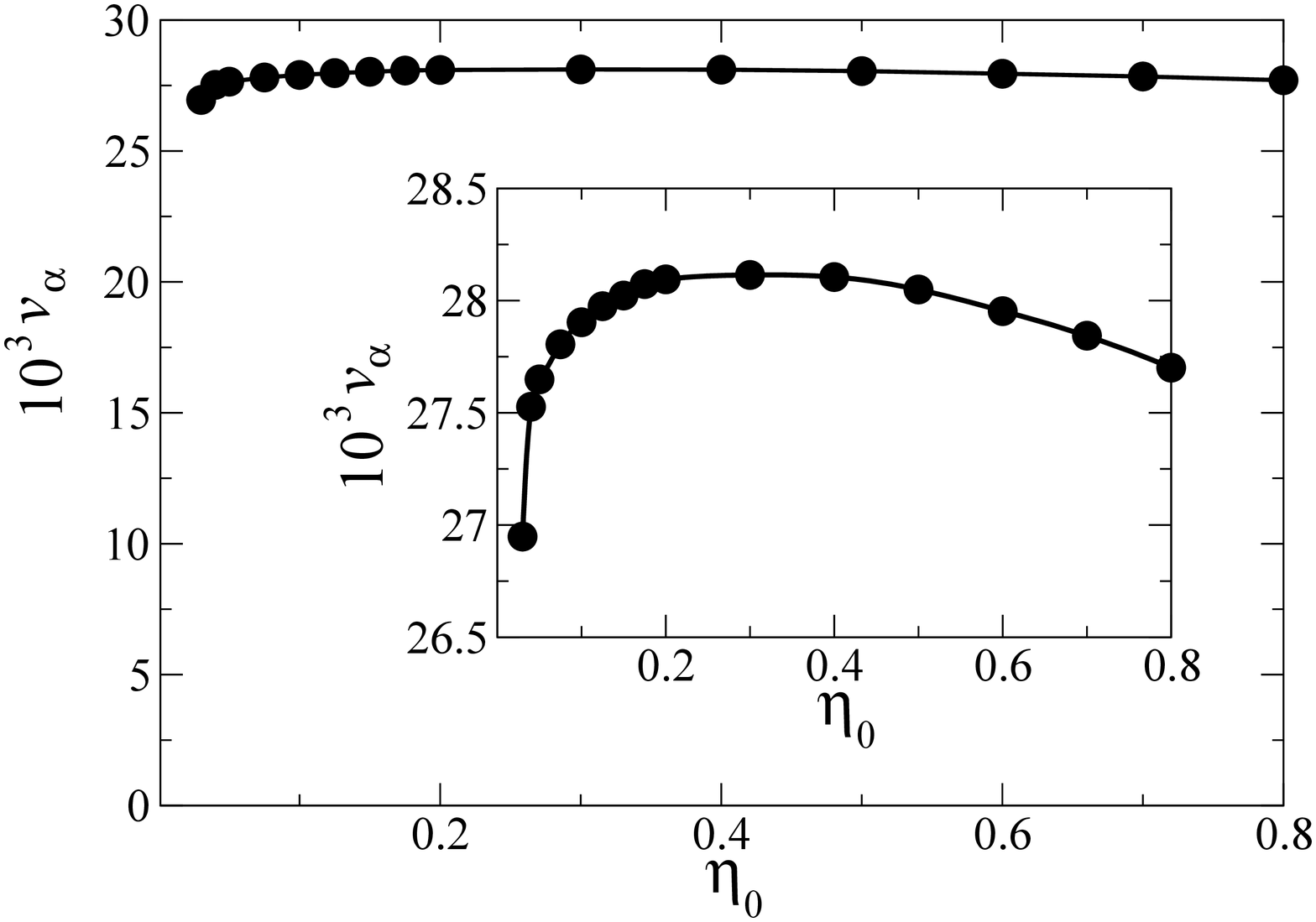}}
\caption{Dependence of maximal $v_\alpha$ in Fig. \ref{Fig4} on $\eta_0$ variations.}
\label{Fig6}       
\end{figure}

The next question we address is whether the subtransport can be further optimized by a variation of 
$\eta_0$ for a maximal value of $v_\alpha$ in Fig. \ref{Fig4}.   The corresponding results are shown in Fig. 
\ref{Fig6}. They reveal that this dependence on $\eta_0$ is rather weak, even if a minor optimization does
take place -- see insert in  Fig. \ref{Fig6}.

\subsection{Load and efficiency}

\begin{figure}
\resizebox{0.75\columnwidth}{!}{\includegraphics{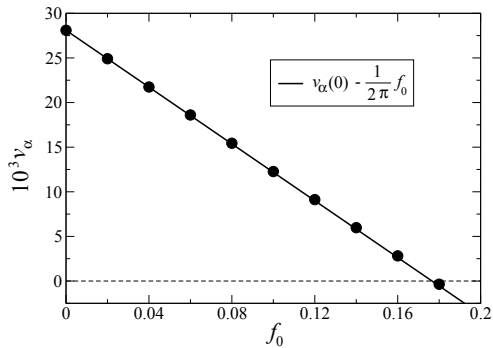}}
\caption{ Dependence of the maximal in Fig. \ref{Fig4} subvelocity $v_\alpha$ on the load $f_0$:
numerics (symbols) \textit{vs.} analytical result (line).}
\label{Fig7}       
\end{figure}

Finally we come to clarifying the question whether the studied anomalous ratchets can do any useful work
and how big can be their thermodynamic efficiency. This question is not a trivial issue at all since
friction can and does play a useful role, contrary to intuition which can mislead. 
Not understanding the essence of the thermal fluctuation-dissipation
theorem (FDT) can drive and mislead research into a wrong direction by tempting to eliminate the dissipative
effects overall and concentrating on the limit of the so-called frictionless or Hamiltonian 
ratchets.  Friction is always associated with dissipative losses and one can believe that
a complete elimination of friction will result in a most efficient motor. However, the FDT
says that at thermal equilibrium the energy dissipated in motion of a Brownian 
particle by friction is regained due to absorption of energy obtained from thermal random forces, so that
the both processes are balanced at the thermal equilibrium, where the total heat exchange between the particle
and its environment is absent. For example, isothermal biological molecular motors can work in spite 
of a strong friction at 
thermodynamic efficiency
close to one (though then infinitely slow, at power close to zero) \cite{Kinosita,Toyabe,Romanovsky}.
 However, subdiffusion introduces new features. Phenomenologically, it can be characterized
by an effectively increasing in time viscous friction. Indeed, 
let us do for a minute an \textit{ad hoc} Markovian
approximation in GLE (\ref{GLE1}) by replacing  $\dot x(t')$ with $\dot x(t)$ in the memory friction
integral (the explicit form of the term formally written with use of the Caputo fractional derivative).
Then, dissipative part of this equation is characterized by an infinitely increasing
in time effective friction $\eta_{\rm eff}(t)=\int_0^t\eta(t')dt'\propto t^{1-\alpha}$. 
This is a simplest way to understand the origin of subdiffusion and subtransport 
in physical terms (though generally one has to be very careful with such an \textit{ad hoc}
approximation, especially in the presence of inertial effects). 
Even if the effective friction increases indefinitely in time, such a ratchet does  
a useful work against a load and
is characterized by a finite stopping force, see in Fig. \ref{Fig7}. Similar was shown also in the presence of
inertial effects, see Fig. 5 in Ref. \cite{GKh12}. The numerics are well described by a simple analytical
dependence 
\begin{equation}\label{v(f)}
v_\alpha(f_0)=v_\alpha(0)-f_0/2\pi,
\end{equation}
which can be inferred by a linear response argumentation given the asymptotical independence of the
subtransport on the presence of periodic potential in the static case, $A\to 0$. The presence of the
factor $1/2\pi$ is due to the used scaling of nondimensional variables. 

Is also the thermodynamic efficiency finite? We define it in a standard way as a portion of input
energy $W_{\rm ext}(t)$ put into a useful work $W_{\rm use}(t)$ done against the loading force
$f_0$, i. e. 
$R(t)=W_{\rm use}(t)/W_{\rm ext}(t)$. To find this quantity we following to the ideas 
presented in \cite{Sekimoto} and
first rewrite Eq. (\ref{GLE1}) as a force balance equation
\begin{equation}\label{force}
-f_{\rm int}(x)=u(t)+f_{\rm ext}(t)-f_0,
\end{equation}
where, $f_{\rm int}(x)=-dU(x)/dx$, is the periodic potential force
acting on the particle (stator potential), and 
\begin{equation}\label{diss_force}
u(t)=-\eta_0\frac{{\rm d} x}{{\rm d}t}-\eta_\alpha\frac{{\rm d}^\alpha x}{{\rm d}t^\alpha}+
\xi_0(t)+\xi_{\alpha}(t)
\end{equation}
is the total stochastic viscoelastic force acting on the particle from the side
of environment. Multiplying the force balance equation
(\ref{force}) by $\dot x(t)$, integrating it within the time interval $[0,t)$, and averaging over a bunch
of trajectories (time averaging is also appropriate since the considered dynamics is ergodic) one obtains the
following energy balance equation 
\begin{equation}\label{energy}
\Delta U(t)=\Delta Q(t)+W_{\rm ext}(t)-W_{\rm use}(t),
\end{equation}
where $\Delta U(t)= U(x(t))-U(x(0))$ is the change of internal energy of the considered Brownian
motor (which is bounded), $\Delta Q(t)=\langle\int_{x(0)}^{x(t)} u(t') dx(t') \rangle$ is the heat exchanged
between the motor particle and its environment, $W_{\rm ext}(t)=\langle\int_{x(0)}^{x(t)} 
f_{\rm ext}(t') dx(t') \rangle$
is the work done by the external force on the whole system, or input energy
provided by it. This input energy is used to do a useful work $W_{\rm use}(t)=f_0\langle x(t)-x(0)\rangle$
against a load.
For a large $t$, the fluctuating change of internal energy is negligible and we have
\begin{equation}\label{energy2}
W_{\rm use}(t)=W_{\rm ext}(t)-|\Delta Q(t)|\;.
\end{equation}
Thermodynamic efficiency is $R(t)=W_{\rm use}(t)/W_{\rm ext}(t)$. Notice that this
 efficiency is zero when
an external loading force is absent. 
Then all the external work done is dissipated as heat absorbed by the environment.
At thermal equilibrium the total heat exchange is absent, $\Delta Q(t)=0$. This is expression of FDT,
which is guaranteed by the FDT condition for GLE. The efficiency of Brownian motors is therefore
maximized when they are operating mostly close to the thermal equilibrium, to minimize the heat losses.
Theoretically, efficiency can reach value of one (if to operate very slowly at almost zero power), 
and some biological molecular
motors can indeed be very efficient, by operating in multiple small steps \cite{Romanovsky,Okada,Inoue}. 
This is a textbook wisdom \cite{NelsonBook}.

\begin{figure}
\resizebox{0.75\columnwidth}{!}{\includegraphics{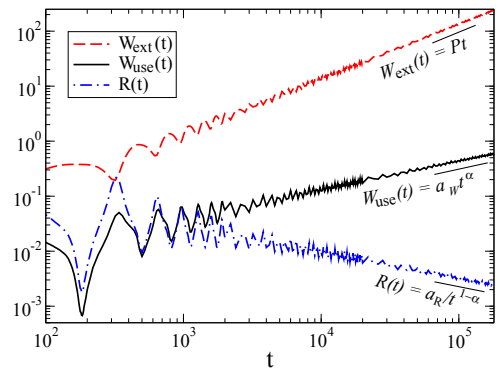}}
\caption{(Color online) Dependencies of the pumped energy, useful work, and thermodynamic efficiency on time for
an optimal load, $\Omega=0.02$, $A=1$, $T=0.075$, and $\eta_0=0.2$.  }
\label{Fig8}       
\end{figure}

The efficiency of subtransport presents new features. Since the transport is subdiffusive, the useful
work scales sublinearly in time, $W_{\rm use}(t)=a_W t^\alpha$, for a sufficiently large $t$.
We can name the coefficient $P_\alpha=a_W \Gamma(1-\alpha)$ subpower, or fractional power. It replaces
for anomalous motors the notion of power, i.e. the work
done per unit of time. The averaged energy pumped by a periodically changing in 
time force scales, however, linearly with time, $W_{\rm ext}(t)=P t$, i.e. with a 
number of oscillations done, 
see in Fig. \ref{Fig8}. It can be characterized by a pumping power $P$. 
For this reason, thermodynamics efficiency decays in time as $R(t)=a_R/t^{1-\alpha}$,
with $a_R=a_W/P$. One can name $a_R$ fractional efficiency.
 Despite this decaying character: (1) The useful work and sub-power are always finite
in the presence of a load bounded by $0<f_0<f_{\rm stop}$, (2) Efficiency decays 
algebraically slowly and compares
well with the efficiency of normal fluctuating tilt ratchets on the time scale of simulations.
In this respect, the efficiency of normal ratchets estimated for the first time in Ref. \cite{Sekimoto}
was as low as $0.01 \%$.
Fluctuating tilt ratchets operate as isothermal Brownian machines rather poor. This is a well known
fact. For larger values of $\alpha$, say for $\alpha \sim 0.9$, considered 
subdiffusive ratchets would operate
with a very slow decaying efficiency, $R(t)\propto 1/t^{0.1}$, on a very long time scale. 
This again confirms that even an infinitely strong  
effective friction (in a naive Markovian approximation assuming a strict subdiffusion)
is not an obstacle for thermodynamical efficiency, paradoxically enough.

\begin{figure}
\resizebox{0.75\columnwidth}{!}{\includegraphics{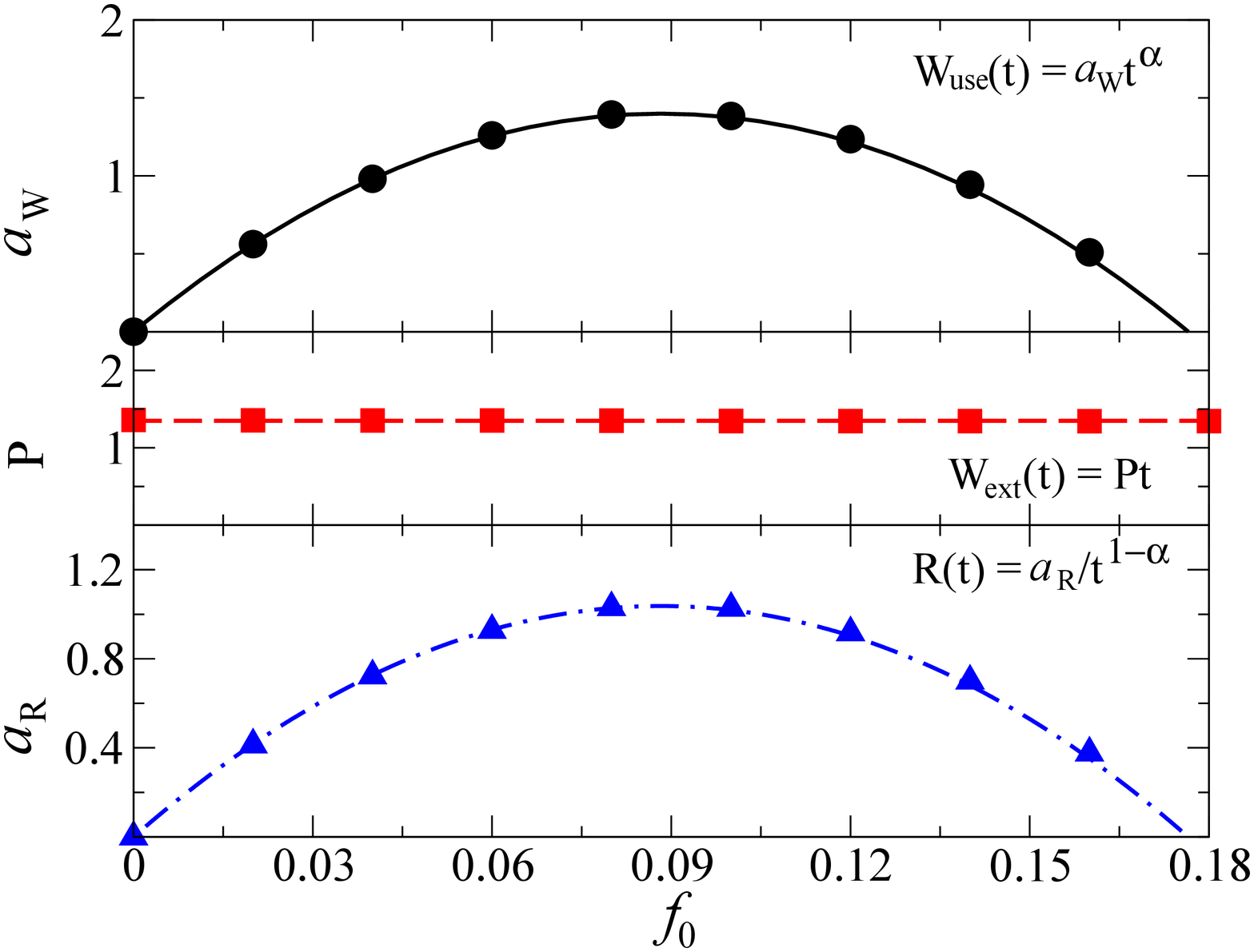}}
\caption{(Color online) Dependencies of the subpower coefficient $a_W$ (in arbitrary units), 
input power $P$ (in arbitrary units), 
and the fractional efficiency
$a_R$ on load. Theoretical results (full lines) agree with numerics (symbols). 
Parameters are same as in Fig. \ref{Fig8}.}
\label{Fig9}       
\end{figure}

Now we are able to provide a very simple theory for subpower coefficient $a_W$ and the fractional 
efficiency coefficient
$a_R$. Clearly, because of $W_{\rm use}(t)=f_0\langle x(t)\rangle 
\sim f_0 v_\alpha(f_0) t^\alpha/\Gamma (1+\alpha)$ and
Eq. (\ref{v(f)}), we have 
\begin{equation}\label{a_W}
a_W(f_0)=f_0[v_\alpha(0)-f_0/2\pi]/\Gamma (1+\alpha)\;.
\end{equation}
This parabolic dependence on $f_0$ agrees with the numerical results in Fig. \ref{Fig9} very well.
Furthermore, since the input power $P$ does not depend on load (the same feature as for normal
diffusion ratchets, see in Fig. \ref{Fig9}), the dependence of fractional efficiency $a_R(f_0)$ 
on load is qualitatively
the same with the only modification, $a_R(f_0)=a_W(f_0)/P$.

\subsubsection{Other definitions of motor efficiency}

Other definitions of the efficiency of Brownian motors were introduced in order to characterize their
performance in the absence of a loading force. Then, all the input energy is dissipated finally 
as heat. The  main idea
is to characterize the motor performance against the dissipative force of the environment when 
a hindering force is absent, or its performance against both environment and hindering force
while motor translocates a cargo. 
Different characterizations have been proposed for normal ratchets \cite{Derenyi,Suzuki,Wang}.
 Let us consider how
they can be modified for anomalous transport and what might be a natural proposal for the efficiency 
in the absence of load, or natural Stokes efficiency. Let us consider the frictional part
of the dissipative force of the environment $u(t)$ in Eq. (\ref{diss_force}).
It is, $u_{\rm diss}(t)=f_{\rm mem}(t)+f_{\rm Stokes}(t)$, or $u_{\rm diss}(t)=\langle u(t)\rangle$.
The definition of the generalized efficiency given in \cite{Derenyi} for normal diffusion
ratchet is
\begin{eqnarray}\label{DBA}
R_{\rm DBA}=\frac{P_{\rm use}+\eta_0\langle v^2\rangle }{P},
\end{eqnarray}
where $P_{\rm use}=\dot W_{\rm use}$ is the useful power, and $\langle v^2\rangle $ is the steady-state averaged 
squared motor velocity. 
$\eta_0\langle v^2\rangle $ is just averaged power of dissipation losses caused by the 
macroscopic Stokes friction $f_{\rm Stokes}$. Another related option proposed is \cite{Suzuki} is to use 
$\langle v\rangle^2 $ instead of $\langle v^2\rangle $, i.e. to neglect the velocity 
fluctuations $\langle \delta v^2\rangle=\langle v^2\rangle-\langle v\rangle^2$ 
in calculating work done against the frictional forces.
The Stokes efficiency is obtained by setting $P_{\rm use}=0$. This corresponds to zero
loading force, $f_0=0$. Though in
\cite{Wang} it was defined differently as 
$R_{\rm Stokes}=\eta_0 \langle v\rangle^2/(P+f_0\langle v\rangle)$.
A generalization of (\ref{DBA}) to the
present case should read
\begin{eqnarray}
R_{\rm DBAgen}(t)=\frac{W_{\rm use}(t)+W_{\rm diss}(t)}{W_{\rm ext}(t)},
\end{eqnarray}
where $W_{\rm diss}(t)=\eta_0\int_0^{t}\langle v^2(t')\rangle dt'+\int_0^{t}dt'
\int_0^{t'} dt''\eta(t'-t'')\langle v(t')v(t'')\rangle $. A different generalization in the
spirit of \cite{Suzuki} would amount to replacing $\langle v^2(t)\rangle$
with  $\langle v(t)\rangle^2$ and  $\langle v(t')v(t'')\rangle$ with
$\langle v(t')\rangle \langle v(t'')\rangle$ in the last expression. We shall
not consider these various generalizations further in the present paper but notice 
that the \textit{total} dissipative force acting from the environment
on the motor particle is yet $u(t)$. It includes also the randomly fluctuating forces.
By the third law of Newton the particle exerts on the environment the force $-u(t)$
and the averaged work done by this force on the environment just equals to the heat
losses $|\Delta Q(t)|$. Therefore, a natural definition for the generalized 
Stokes efficiency
would be just $R_{\rm Stokes}(t)=|\Delta Q(t)|/W_{\rm ext}(t)=1-R(t)$. It equals 100 \% in the
absence of load and approaches this maximal value even in the presence of load asymptotically
for the studied anomalous ratchets. In other words, the work on translocation
of a cargo in such a highly dissipative viscoelastic environment is done mostly against 
resistance of this
environment. This is a natural conclusion.

\section{Summary and conclusions}

We studied a model of overdamped subdiffusive ratchets rocked by a
time-periodic force in viscoelastic media characterized by a power-law decaying memory kernel.
The Brownian motor particle is subjected to viscous friction and white thermal noise.
Viscoelastic medium's degrees of freedom were modeled by auxiliary Brownian
particles that are elastically coupled to the central Brownian particle. 
Some of these auxiliary Brownian particles  are extremely slow. They imprint
memory about former positions of the central Brownian particle and create a retarding viscoelastic 
force causing anomalous diffusion and transport. 
Just a handful of such auxiliary Brownian particles suffices to
model subdiffusion and subtransport on practically any experimentally relevant time scale.
Our setup is fully equivalent to a Generalized Langevin Equation, where
the memory kernel and random force of environment are related by the
fluctuation-dissipation relation at ambient temperature of the
environment. The setup of our modeling is ergodic
and the Brownian motor subvelocity can be found from a single particle trajectory,
though an ensemble averaging over $10^4$ particles has been done to obtain most results presented. 
Subdiffusive current is optimized with the frequency of periodical driving
and temperature. It depends also on the viscous friction acting directly on the motor particle. 
However, the subdiffusion
coefficient depends weakly on other parameters being linearly proportional  to temperature
within the model of temperature-independent friction. The directed ratchet subtransport
can possess  thus at low temperature a very good quality, as characterized
by the generalized Peclet number ${\rm Pe}_\alpha$. 

Furthermore, we have shown that the considered anomalous Brownian motors 
are able to sustain a substantial load and do a useful work which scales sublinearly
with time and can be characterized by subpower. Since the energy pumped by the external
time-periodic force scales linearly with time the motor efficiency decays algebraically
in time. It can, however, favorably agree with the efficiency of the normal Brownian motors
of the kind considered on an appreciably long time intervals. We provided a simple theory
for thermodynamic efficiency of anomalous Brownian motors which agrees remarkably well
with the numerical results obtained.

We expect that nontrivial results obtained in this work will 
stimulate a further cross-fertilization between the fields of anomalous
diffusion and transport and the field of fluctuation-induced transport in the absence
of a biasing on average force. A further generalization of the model presented here towards
flashing potential  ratchets  opens a way
to treat operation of molecular motors in such viscoelastic environments as cytosol of biological
cells. The corresponding work is in progress.

\section*{Acknowledgment} 
Support of this research by the Deutsche Forschungsgemeinschaft (German Research Foundation), Grants
GO 2052/1-1 and GO 2052/1-2 is gratefully acknowledged.

\end{document}